\documentclass{emulateapj}

\shorttitle{EVOLVING TO SNe Ia WITH SHORT DELAY TIMES}
\shortauthors{WANG et al.}

\begin{document}
\title{EVOLVING TO TYPE Ia SUPERNOVAE WITH SHORT DELAY TIMES}

\author{Bo Wang,\altaffilmark{1,2} Xuefei Chen,\altaffilmark{1} Xiangcun Meng,\altaffilmark{3} and Zhanwen Han\altaffilmark{1}}
\altaffiltext{1}{National Astronomical Observatories/Yunnan
Observatory, the Chinese Academy of Sciences, Kunming 650011, China;
wangbo@ynao.ac.cn, zhanwenhan@ynao.ac.cn} \altaffiltext{2}{Graduate
University of Chinese Academy of Sciences, Beijing 100049, China}
\altaffiltext{3}{Department of Physics and Chemistry, Henan
Polytechnic University, Jiaozuo 454003, China}

\begin{abstract}\label{abstract}
The single-degenerate model is currently a favourable progenitor
model for Type Ia supernovae (SNe Ia). Recent investigations on the
WD + He star channel of the single-degenerate model imply that this
channel is noteworthy for producing SNe Ia. In this paper we studied
SN Ia birthrates and delay times of this channel via a detailed
binary population synthesis approach. We found that the Galactic SN
Ia birthrate from the WD + He star channel is $\sim 0.3\times
10^{-3}\ {\rm yr}^{-1}$ according to our standard model, and that
this channel can explain SNe Ia with short delay times
($\sim4.5\times10^7 - 1.4\times10^8$\,yr). Meanwhile, these WD + He
star systems may be related to the young supersoft X-ray sources
prior to SN Ia explosions.

\end{abstract}
\keywords{binaries: close --- stars: evolution --- supernovae:
general --- white dwarfs}

\section{INTRODUCTION}\label{1:INTRODUCTION}
Type Ia supernovae (SNe Ia) play an important role in astrophysics,
especially in the study of cosmic evolution. They have been applied
successfully in determining cosmological parameters (e.g., $\Omega$
and $\Lambda$; Riess et al. 1998; Perlmutter et al. 1999). It is
widely accepted that SNe Ia are thermonuclear explosions of
carbon-oxygen (CO) white dwarfs (WDs) accreting matter from their
companions (for a review see Nomoto et al. 1997). However, there is
still no agreement on the nature of their progenitors (Hillebrandt
\& Niemeyer 2000; R\"{o}pke \& Hillebrandt 2005; Wang et al. 2008;
Podsiadlowski et al. 2008), and this may raise doubts about the
distance calibration which is purely empirical and based on the SN Ia
sample of the low red-shift Universe.

At present, various progenitor models of SNe Ia can be examined by
comparing the distribution of the delay time (between the star
formation and SN Ia explosion) expected from a progenitor channel
with that of observations (e.g., Chen $\&$ Li 2007; Xu $\&$ Li 2009;
L\"{u} et al. 2009; Mannucci 2009; Schawinski 2009). Recently, there
are three important observational results for SNe Ia, i.e., the
strong enhancement of the SN Ia birthrate in radio-loud early-type
galaxies, the strong dependence of the SN Ia birthrate on the colors
of the host galaxies, and the evolution of the SN Ia birthrate with
redshift (Della Valle et al. 2005; Mannucci et al. 2005, 2006,
2008). The relation between SN Ia birthrate and radio power
implicates the information on the time-scales of the order of
$10^{8}$\,yr, which corresponds to the radio activity lifetime; the
strong dependence of the local birthrate on the colors of the host
galaxies is related to the time-scales of the order of the galaxy
color evolution (i.e., $0.5-1$\,Gyr); the evolution of the SN Ia
birthrate with redshift is sensitive to the long time-scales (a few
Gyr) (for details see Mannucci et al. 2006). According to the
present observational results, Mannucci et al. (2006) found that
they can be best matched by a bimodal delay time distribution, in
which about half of the SNe Ia explode soon after starburst, with a
delay time less than $\sim 10^8$\,yr, while those remaining have a
much wider distribution, which can be well described by an
exponential function with a decay time of about 3\,Gyr (see also
Mannucci 2008). Note that Scannapieco \& Bildsten (2005) explored
the two components of SN Ia birthrates and found that a young SN Ia
population may be helpful to explain the Fe content of the
intracluster medium in galaxy clusters. Moreover, by investigating
the star formation history of 257 SN Ia host galaxies, Aubourg et
al. (2008) recently found evidence of a short-lived population of SN
Ia progenitors with lifetimes of less than 180\,Myr.

Over the last decades, two competing progenitor models of SNe Ia
were discussed frequently, i.e., the single-degenerate (SD) and
double-degenerate (DD) models. Of these two progenitor models, the
SD model (Whelan \& Iben 1973; Nomoto et al. 1984; Fedorova et al.
2004; Han 2008; Meng et al. 2009) is widely accepted at present. It
is suggested that the DD model, which involves the merger of two CO
WDs (Iben \& Tutukov 1984; Webbink 1984; Han 1998), likely leads to
an accretion-induced collapse rather than a SN Ia (Nomoto \& Iben
1985; Saio \& Nomoto 1985; Timmes et al. 1994). For the SD model,
the companion is probably a MS star or a slightly evolved subgiant
star (WD + MS channel), or a red-giant star (WD + RG channel).
However, these two SD channels did not predict such young SN Ia
populations (Hachisu et al. 1996, 1999a, 1999b; Li $\&$ van den
Heuvel 1997; Langer 2000; Han $\&$ Podsiadlowski 2004,
2006).\footnote{Note that Hachisu et al. (2008) investigated new
evolutionary models for SN Ia progenitors, introducing the
mass-stripping effect on a MS or slightly evolved companion star by
winds from a mass-accreting WD. The model can explain the presence
of very young ($\la 10^{8}$\,yr) populations of SN Ia progenitors,
but the model depends on the efficiency of the mass-stripping
effect.}

Wang et al. (2009) recently studied a WD + He star channel for the
SD model to produce SNe Ia. In the study they carried out detailed
binary evolution calculations of this channel for about 2600 close
WD binaries with metallicity $Z=0.02$, in which a CO WD accretes
material from a He MS star or a He subgiant to increase its mass to
the Chandrasekhar-mass limit. The study showed the SN Ia production
regions in the ($\log P^{\rm i}, M^{\rm i}_2$) plane (see Fig. 8 of
Wang et al. 2009), where $P^{\rm i}$ and $M^{\rm i}_2$ are the
orbital period and the mass of the He companion star at the onset of
the Roche lobe overflow (RLOF), respectively, and indicated that
this channel is noteworthy for producing SNe Ia. Because the WD + He
star systems are from intermediate mass binary systems, this channel
is likely to explain SNe Ia with short delay times. However, SN Ia
birthrates and delay times through this channel are not well known
from a viewpoint of the binary population synthesis (BPS).

The purpose of this paper is to study SN Ia birthrates for the WD +
He star channel and to explore possible SN Ia progenitor systems
with short delay times from this channel. In Section 2, we describe
the BPS approach for the WD + He star channel. The simulation
results of the BPS approach is shown in Section 3. Finally,
discussion and conclusion are given in Section 4.

\section{BINARY POPULATION SYNTHESIS}\label{SECT:BINARY POPULATION SYNTHESIS}
In order to investigate SN Ia birthrates and delay times for the WD
+ He star channel, we have performed a series of Monte Carlo
simulations in the BPS study. In each simulation, by using the
Hurley's rapid binary evolution code (Hurley et al. 2000, 2002), we
have followed the evolution of $4\times10^{\rm 7}$ sample binaries
from the star formation to the formation of the WD + He star systems
according to three evolutionary channels (Sect. 2.2). We assumed
that, if the parameters of a CO WD + He star system at the onset of
the RLOF are located in the SN Ia production regions in the ($\log
P^{\rm i}, M^{\rm i}_2$) plane (Fig. 8 of Wang et al. 2009), a SN Ia
is produced. Hereafter, we use the term {\sl primordial} to
represent the binaries before the formation of WD + He star systems.

\subsection{Common Envelope in Binary Evolution}
When the primordial primary (massive star) in a binary system fills
its Roche lobe, the primordial mass ratio (primary to secondary) is
crucial for the mass transfer. If it is larger than a critical mass
ratio, $q_{\rm c}$, the mass transfer may be dynamically unstable
and a common envelope (CE) forms (Paczy\'{n}ski 1976). The mass
ratio $q_{\rm c}$ varies with the evolutionary state of the
primordial primary at the onset of RLOF (Hjellming \& Webbink 1987;
Webbink 1988; Han et al. 2002; Podsiadlowski et al. 2002). In this
study we adopt $q_{\rm c}$ = 4.0 when the primary is in the MS stage
or Hertzsprung gap. This value is supported by detailed binary
evolution studies (Han et al. 2000; Chen \& Han 2002, 2003). If the
primordial primary is on the first giant branch (FGB) or asymptotic
giant branch (AGB) stage, we use
\begin{equation}
q_{\rm c}=[1.67-x+2(\frac{M_{\rm c1}^{\rm P}}{M_{\rm 1}^{\rm
P}})^{\rm 5}]/2.13,
  \end{equation}
where $M_{\rm 1}^{\rm P}$ is the mass of the primordial primary,
$M_{\rm c1}^{\rm P}$ is the core mass of the primordial primary, and
$x={\rm d}\ln R_{\rm 1}^{\rm P}/{\rm d}\ln M_{\rm 1}^{\rm p}$ is the
mass-radius exponent of the primordial primary and varies with
composition. If the mass donor stars (primaries) are naked He
giants, $q_{\rm c}$ = 0.748 based on equation (1) (see Hurley et al.
2002 for details).

When a CE forms, the embedded in the CE is a `new' binary consisting
of the dense core of the primordial primary and the primordial
secondary. Owing to frictional drag within the envelope, the orbit
of the `new' binary decays and a large part of the orbital energy
released in the spiral-in process is injected into the envelope
(Livio \& Soker 1988). The CE ejection is still an open problem.
Here, we use the standard energy equations (Webbink 1984) to
calculate the output of the CE phase. The CE is ejected if
\begin{equation}
 \alpha_{\rm ce} \left( {G M_{\rm don}^{\rm f} M_{\rm acc} \over 2 a_{\rm f}}
- {G M_{\rm don}^{\rm i} M_{\rm acc} \over 2 a_{\rm i}} \right) = {G
M_{\rm don}^{\rm i} M_{\rm env} \over \lambda R_{\rm don}},
\end{equation}
where $\lambda$ is a structure parameter that depends on the
evolutionary stage of the donor, $M_{\rm don}$ is the mass of the
donor, $M_{\rm acc}$ is the mass of the accretor, $a$ is the orbital
separation, $M_{\rm env}$ is the mass of the donor's envelope,
$R_{\rm don}$ is the radius of the donor, and the indices ${\rm i}$
and ${\rm f}$ denote the initial and final values, respectively. The
right side of the equation represents the binding energy of the CE,
the left side shows the difference between the final and initial
orbital energy, and $\alpha_{\rm ce}$ is the CE ejection efficiency,
i.e., the fraction of the released orbital energy used to eject the
CE. For this prescription of the CE ejection, there are two highly
uncertain parameters (i.e., $\lambda$ and $\alpha_{\rm ce}$).  We
usually set $\lambda$ to be 0.5 to constrain $\alpha_{\rm ce}$ (de
Kool 1990), although an exact calculation should take into account
the issue that $\lambda$ depends on the stellar structure. In
principle, we expect $0<\alpha_{\rm ce}\leq1$, but we often find
that $\alpha_{\rm ce}$ exceeds 1 for the purpose of explaining
observed binaries. This may indicate that other energy sources may
also contribute to the ejection of the envelope, e.g., the internal
energy of the envelope (Han et al. 1994, 1995; Podsiadlowski et al.
2003; Webbink 2008). As in previous studies, we combine $\alpha_{\rm
ce}$ and $\lambda$ into one free parameter $\alpha_{\rm ce}\lambda$,
and set it to be 0.5 and 1.5 (e.g., L\"{u} et al. 2006).

\subsection{Evolutionary Channels to WD + He Star Systems}
According to the evolutionary phase of the primordial primary at the
beginning of the first RLOF, there are three channels which can
produce CO WD + He star systems and then produce SNe Ia.

(1) {\em He star channel.} The primordial primary first fills its
Roche lobe when it is in the subgiant or RG stage (Case B mass
transfer defined by Kippenhahn \& Weigert 1967). At the end of the
RLOF, the primary becomes a He star and continues to evolve. After
the exhaustion of central He, the He star which now contains a CO
core may fill its Roche lobe again due to expansion of the He star
itself, and transfer its remaining He-rich envelope to the MS
companion star, eventually leading to the formation of a CO WD + MS
system. After that, the MS companion star continues to evolve and
fills its Roche lobe in the subgiant or RG stage. A CE is possibly
formed quickly because of dynamically unstable mass transfer. If the
CE can be ejected, a close CO WD + He star system is then produced.
The CO WD + He star system continues to evolve, and the He star may
fill its Roche lobe again (due to orbit decay induced by the
gravitational wave radiation or the expansion of the He star
itself), and transfer some material onto the surface of the CO WD.
The accreted He may be converted into C and O via He-shell burning,
and the CO WD increases in mass and explodes as a SN Ia when its
mass reaches the Chandrasekhar mass limit. For this channel, SN Ia
explosions occur for the ranges $M_{\rm 1,i}\sim5.0-8.0\,M_\odot$,
$M_{\rm 2,i}\sim2.0-6.5\,M_\odot$ and $P^{\rm i} \sim 10-40$\,days,
where $M_{\rm 1,i}$, $M_{\rm 2,i}$ and $P^{\rm i}$ are the initial
mass of the primary and the secondary at ZAMS, and the initial
orbital period of a binary system.

(2) {\em EAGB channel.} If the primordial primary is on the early
AGB (EAGB, i.e., He is exhausted in the center of the star while
thermal pulses have not yet started), a CE will be formed because of
dynamically unstable mass transfer. After the CE is ejected, the
orbit decays and the primordial primary becomes a He RG. The He RG
may fill its Roche lobe and start mass transfer, which is likely
stable and leaves a CO WD + MS system. The following evolution of
the CO WD + MS system is similar to that in the {\em He star
channel} above, and may form a CO WD + He star system and finally
produce a SN Ia. For this channel, SN Ia explosions occur for the
ranges $M_{\rm 1,i}\sim6.0-6.5\,M_\odot$, $M_{\rm
2,i}\sim5.5-6.0\,M_\odot$ and $P^{\rm i} \sim 300-1000$\,days.

(3) {\em TPAGB channel.} The primordial primary fills its Roche lobe
at the thermal pulsing AGB (TPAGB) stage, and the companion star
evolves to a He-core burning stage. A CE is easily formed owing to
dynamically unstable mass transfer during the RLOF. After the CE
ejection, the primordial primary becomes a CO WD, then a CO WD + He
star system is produced. The following evolution of the CO WD + He
star system is similar to that in two channels above, i.e., a SN Ia
may be produced finally. For this channel, SN Ia explosions occur
for the ranges $M_{\rm 1,i}\sim5.5-6.5\,M_\odot$, $M_{\rm
2,i}\sim5.0-6.0\,M_\odot$ and $P^{\rm i}\ga 1000$\,days.

\subsection{Basic Parameters for Monte Carlo Simulations}

In the BPS study, the Monte Carlo simulation requires as input the
initial mass function (IMF) of the primary, the mass-ratio
distribution, the distribution of initial orbital separations, the
eccentricity distribution of binary orbit and the star formation
rate (SFR).

(1) The IMF of Miller \& Scalo (1979, MS79) is adopted. The
primordial primary is generated according to the formula of Eggleton
et al. (1989)
\begin{equation}
M_{\rm 1}^{\rm p}=\frac{0.19X}{(1-X)^{\rm 0.75}+0.032(1-X)^{\rm
0.25}},
  \end{equation}
where $X$ is a random number uniformly distributed in the range [0,
1] and $M_{\rm 1}^{\rm p}$ is the mass of the primordial primary,
which ranges from 0.1\,$M_{\rm \odot}$ to 100\,$M_{\rm \odot}$. The
studies of the IMF by Kroupa et al. (1993) and Zoccali et al. (2000)
support this IMF. As an alternative IMF we also consider the IMF of
Scalo (1986, S86)
\begin{equation} M_{\rm 1}^{\rm
p}=0.3\left(\frac{X}{1-X}\right)^{0.55},
  \end{equation}
where the meanings of $X$ and $M_{\rm 1}^{\rm p}$  are similar to
that of equation (3).

(2) The initial mass-ratio distribution of the binaries, $q'$, is
quite uncertain for binary evolution. For simplicity, we take a
constant mass-ratio distribution (Mazeh et al. 1992; Goldberg \&
Mazeh 1994):
\begin{equation}
n(q')=1, \hspace{2.cm} 0<q'\leq1,
\end{equation}
where $q'=M_{\rm 2}^{\rm p}/M_{\rm 1}^{\rm p}$. This constant
mass-ratio distribution is supported by the study of Shatsky \&
Tokovinin (2002). As alternatives we also consider a rising mass
ratio distribution
\begin{equation}
n(q')=2q',\qquad  0\leq q' \leq 1,
\end{equation}
and the case where both binary components are chosen randomly and
independently from the same IMF (uncorrelated).

(3) We assume that all stars are members of binary systems and that
the distribution of separations is constant in $\log a$ for wide
binaries, where $a$ is separation and falls off smoothly at small
separation:
\begin{equation}
a\cdot n(a)=\left\{
 \begin{array}{lc}
 \alpha_{\rm sep}(a/a_{\rm 0})^{\rm m}, & a\leq a_{\rm 0},\\
\alpha_{\rm sep}, & a_{\rm 0}<a<a_{\rm 1},\\
\end{array}\right.
\end{equation}
where $\alpha_{\rm sep}\approx0.07$, $a_{\rm 0}=10\,R_{\odot}$,
$a_{\rm 1}=5.75\times 10^{\rm 6}\,R_{\odot}=0.13\,{\rm pc}$ and
$m\approx1.2$. This distribution implies that the numbers of wide
binary systems per logarithmic interval are equal, and that about
50\,percent of stellar systems have orbital periods less than
100\,yr (Han et al. 1995).

(4) A circular orbit is assumed for all binaries. The orbits of
semidetached binaries are generally circularized by the tidal force
on a timescale which is much smaller than the nuclear timescale.
Moreover, a binary is expected to become circularized during the
RLOF. As an alternative, we also consider a uniform eccentricity
distribution in the range [0, 1].

(5) We simply assume a constant SFR over the last 15\,Gyr or,
alternatively, as a delta function, i.e., a single starburst. In the
case of a constant SFR, we assume that a binary calibrated with its
primary more massive than $0.8\,M_{\odot}$ is formed annually (see
Iben \& Tutukov 1984; Han et al. 1995; Hurley et al. 2002). From
this calibration, we can get ${\rm SFR}=5\,M_{\rm \odot}{\rm
yr}^{-1}$ (see also Willems \& Kolb 2004). For the case of a single
starburst, we assume a burst producing $10^{11}\,M_{\odot}$ in
stars. In fact, the SFR in a galaxy is neither a constant nor a
delta function over the last 15\,Gyr. A galaxy may have a
complicated star formation history. We only choose these two
extremes for a simplicity. A constant SFR is similar to the
situation of our Galaxy (Yungelson \& Livio 1998; Han $\&$
Podsiadlowski 2004), while a delta function to the situation of
elliptical galaxies or globular clusters. Under the assumption of
the SFR as a delta function, one can obtain the delay time of SNe Ia
from a progenitor channel, and then to compare with that of
observations (e.g., Han $\&$ Podsiadlowski 2004).

\section{THE RESULTS OF BINARY POPULATION SYNTHESIS}

\subsection{Birthrates of SNe Ia}

\begin{table}
 \begin{minipage}{85mm}
 \caption{Galactic birthrates of SNe Ia for different simulation sets, where set 1 is our standard model. $\alpha_{\rm ce}\lambda$ = CE ejection parameter;
$n(q')$ = initial mass ratio distribution; ${\rm IMF}$ = initial
mass function; ${\rm ecc}$ = eccentricity distribution of binary
orbit; $\nu$ = Galactic birthrates of SNe Ia.}
   \begin{tabular}{cccccc}
\hline \hline
Set & $\alpha_{\rm ce}\lambda$ & $n(q')$ & ${\rm IMF}$ & ${\rm ecc}$ & $\nu$ ($10^{-3}$\,yr$^{-1}$)\\
\hline
$1$ & $0.5$ & ${\rm Constant}$     & ${\rm MS79}$    & ${\rm Circular}$ & $0.295$\\
$2$ & $1.5$ & ${\rm Constant}$     & ${\rm MS79}$    & ${\rm Circular}$ & $0.302$\\
$3$ & $1.5$ & ${\rm Constant}$     & ${\rm MS79}$    & ${\rm Uniform}$  & $0.284$\\
$4$ & $1.5$ & ${\rm Constant}$     & ${\rm S86}$     & ${\rm Circular}$ & $0.168$\\
$5$ & $1.5$ & ${\rm Rising}$       & ${\rm MS79}$    & ${\rm Circular}$ & $0.232$\\
$6$ & $1.5$ & ${\rm Uncorrelated}$ & ${\rm MS79}$    & ${\rm Circular}$ & $0.040$\\
\hline
\end{tabular}
\end{minipage}
\end{table}

We performed six sets of simulations (see Table 1) with metallicity
$Z=0.02$ to systematically investigate Galactic birthrates of SNe Ia
for the WD + He star channel, where set 1 is our standard model with
the best choice of model parameters (e.g., Han et al. 2002, 2003,
2007). We vary the model parameters in the other sets to examine
their influences on the final results.

\begin{figure}
\includegraphics[width=6.cm,angle=270]{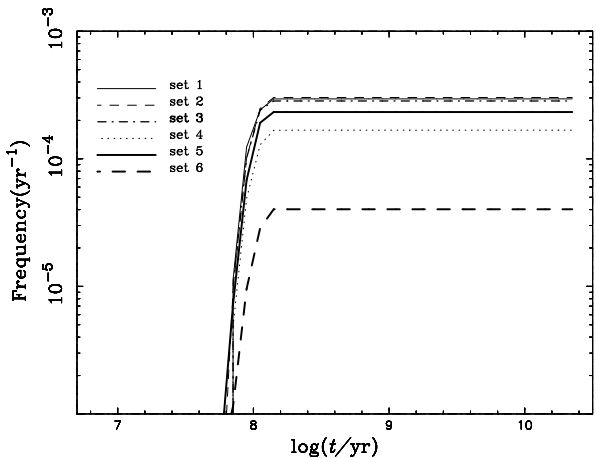}
 \caption{The evolution of Galactic birthrates of SNe Ia for a constant
star formation rate ($Z=0.02$, ${\rm SFR}=5\,M_{\rm \odot}{\rm
yr}^{-1}$). The key to the line-styles representing different sets
is given in the upper left corner. The results of sets 2 and 3
almost coincide with that of set 1.}
\end{figure}

In Figure 1, we show Galactic birthrates of SNe Ia for the WD + He
star channel by adopting $Z=0.02$ and ${\rm SFR}=5\,M_{\rm
\odot}{\rm yr}^{-1}$. The simulation for our standard model (set 1)
gives Galactic SN Ia birthrate of $\sim 0.3\times 10^{-3}\ {\rm
yr}^{-1}$, which is lower than that inferred observationally (i.e.,
$3 - 4\times 10^{-3}\ {\rm yr}^{-1}$; van den Bergh \& Tammann 1991;
Cappellaro \& Turatto 1997). This implies that the WD + He star
channel is only a subclass of SN Ia production, and there may be
some other channels or mechanisms also contributing to SNe Ia, e.g.,
WD + MS channel, WD + RG channel or double-degenerate channel (see
Meng et al. 2009 for details). Especially, as mentioned by Han \&
Podsiadlowski (2004), the WD + MS channel can give a Galactic
birthrate of $\sim0.6-0.8\times 10^{-3}\ {\rm yr}^{-1}$, and is
considered to be an important channel to produce SNe Ia.

According to the results of the six sets of Monte Carlo simulations,
we find that the BPS is sensitive to uncertainties in some input
parameters, in particular the mass-ratio distribution. If we adopt a
mass-ratio distribution for un-correlated component masses (set 6),
the birthrate will decrease to be $\sim 4\times10^{-5}\,{\rm
yr}^{-1}$, as most of the donors in the WD + He star channel are not
very massive which has the consequence that WDs cannot accrete
enough mass to reach the Chandrasekhar-mass limit.

The SN Ia birthrate in galaxies is the convolution of the
distribution of the delay times (DDT) with the star formation
history (SFH) (e.g., Greggio et al. 2008):
\begin{equation}
\nu(t)=\int^t_0 SFR(t-t')DDT(t')dt',
\end{equation}
where the $SFR$ is the star formation rate, and $t'$ is the delay
times of SNe Ia. Due to a constant SFR adopted in this paper, the SN
Ia birthrate $\nu(t)$ is only related to the $DDT$, which can be
expressed by
\begin{equation}
DDT(t)=\left\{
\begin{array}{lc}
0, & t<{\rm t_1},\\
DDT'(t) , &   {\rm t_1} \leq t \leq{\rm t_2},\\
0, & t>{\rm t_2},\\
\end{array}\right.
\end{equation}
where ${\rm t_1}$ and ${\rm t_2}$ are the minimum and maximum delay
times of SNe Ia, respectively, and the $DDT'$ is the distribution of
the delay times between ${\rm t_1}$ and ${\rm t_2}$. When $t$ is
larger than the ${\rm t_2}$, the equation (8) can be written as
\begin{equation}
\nu(t)={\rm SFR}\int^{\rm t_2}_{\rm t_1}DDT'(t')dt'={\rm constant}.
\end{equation}
Therefore, the SN Ia birthrates shown in figure 1 seems to be so
completely flat after the first rise.

\begin{figure}
\includegraphics[width=6.cm,angle=270]{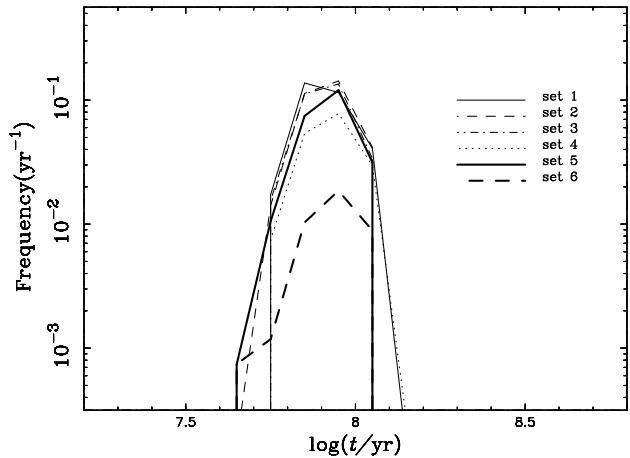}
 \caption{Similar to Fig. 1, but for a single
starburst with a total mass of $10^{\rm 11}M_{\odot}$. The result of
set 3 almost coincides with that of set 2.}
\end{figure}

Figure 2 displays the evolution of SN Ia birthrates for a single
starburst with a total mass of $10^{11}\,M_{\odot}$. In the figure
we see that SN Ia explosions occur between $\sim4.5\times10^7$\,yr
and $\sim1.4\times10^8$\,yr after the starburst, which can explain
SNe Ia with short delay times (Scannapieco \& Bildsten 2005;
Mannucci et al. 2006; Aubourg et al. 2008). The minimum delay time
in the figure is mainly decided by the MS lifetime of a
8$\,M_{\odot}$ star (it is also the maximum mass for the progenitors
of CO WDs). Moreover, after the primordial binary system evolves to
a WD + He star system, the MS lifetime of the He companion star also
contributes to the minimum time, but the time is short, e.g., the MS
lifetime of a 1$\,M_{\odot}$ He star is only about 15\,Myr (Eggleton
2006).

\subsection{Distribution of Initial Parameters of WD + He Star Systems for SNe Ia}

\begin{figure}
\includegraphics[width=9.0cm,angle=0]{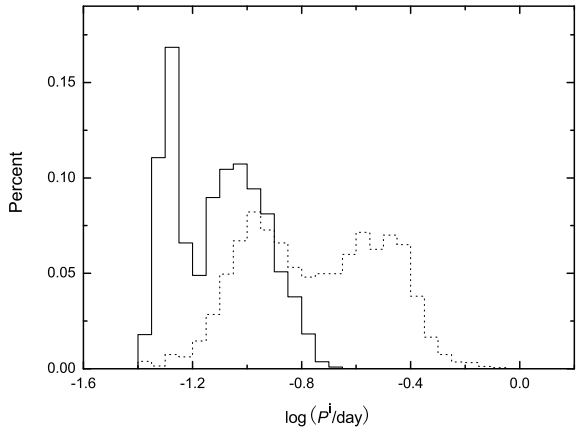}
 \caption{The distribution of the initial orbital periods of the WD +
 He star systems which can ultimately produce SNe Ia for different
 $\alpha_{\rm ce}\lambda$. The simulation uses a metallicity
 $Z=0.02$ and a constant initial
mass-ratio distribution. The solid and the dotted histograms
represent the cases with $\alpha_{\rm ce}\lambda=0.5$ (set 1) and
$\alpha_{\rm ce}\lambda=1.5$ (set 2), respectively. The number in
every case is normalized to 1.}
\end{figure}

\begin{figure}
\includegraphics[width=9.0cm,angle=0]{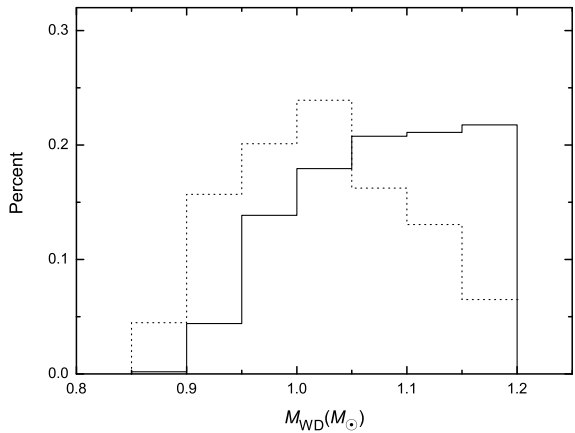}
 \caption{Similar to Fig. 3, but for the distribution of the initial masses of the CO WDs.}
\end{figure}

\begin{figure}
\includegraphics[width=9.0cm,angle=0]{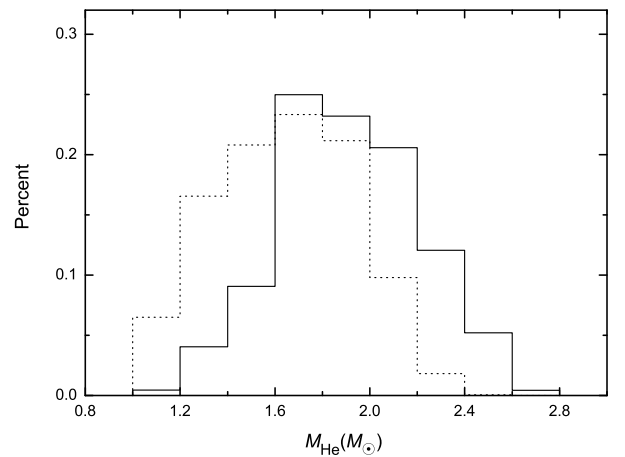}
 \caption{Similar to Fig. 3, but for the distribution of the initial masses of the He donor stars.}
\end{figure}

Observationally, some WD + He star systems are possible SN Ia
progenitors (Wang et al. 2009). Further investigations are necessary
for final confirmation of this (from both observations and
theories). In this section, we will present some properties of
initial WD + He star systems for SNe Ia according to our BPS
approach, which may help to search for potential SN Ia progenitors.

Figure 3 shows the distribution of the initial orbital periods of
the WD + He star systems that ultimately produce SNe Ia with
different $\alpha_{\rm ce}\lambda$. The simulation uses a
metallicity $Z=0.02$ and a constant initial mass-ratio distribution.
The figure displays a result of the current epoch for a constant
SFR. In the figure we can see that there are obviously two peaks for
each case. The left peak in these two cases results from the {\em He
star channel}. Many of the SNe Ia in the right peak are also from
the {\em He star channel}, while others from the {\em EAGB channel}
and the {\em TPAGB channel} (see Sect. 2.2). The {\em He star
channel} has an important contribution to the formation of SNe Ia.
This figure also shows that a high value of $\alpha_{\rm ce}\lambda$
leads to wider WD binaries, since a high value of $\alpha_{\rm
ce}\lambda$ is easier to eject the CE in the binary evolution.

Figure 4 represents the distribution of the initial masses of the CO
WDs. In the figure, a low value of $\alpha_{\rm ce}\lambda$ tends to
have larger WD masses on average. The {\em He star channel} in Sect.
2.2, which allows stable RLOF to produce massive WDs (rather lead to
dynamical mass transfer and a CE phase), is useful to understand
this trend. According to our BPS simulations, we find that a low
value of $\alpha_{\rm ce}\lambda$ will produce more SNe Ia through
the {\em He star channel} than other two channels, and then produce
more massive WDs on average. Figure 5 displays the distribution of
the initial masses of the He donor stars. A low value of
$\alpha_{\rm ce}\lambda$ in the figure tends to have larger He star
masses on average. This is also related to the stable RLOF, which
leads to more massive companion star resulting in the final larger
He-core mass (He star mass). Moreover, a massive He donor star in
the WD + He star channel will evolve more quickly and hence produce
a SN at an earlier time.

When the WDs in Figure 4 increase their masses to the
Chandrasekhar-mass limit, they will explode as SNe Ia. Meanwhile,
the He donor stars in Figure 5, which afford material to the WDs
though the RLOF, will loss a significant amount of masses. We find
that the He stars have masses $\sim0.6-1.7\,M_{\odot}$ at the moment
of SN explosions. Marietta et al. (2000) presented several
high-resolution two-dimensional numerical simulations of the impacts
of SN Ia explosions with companions. The impact makes the companion
in the WD + MS channel lose a mass of $0.15-0.17\,M_\odot$, but the
impact in the WD + He star channel is still unknown. The companion
in the WD + He star channel may lose more masses than that of the WD
+ MS channel. This is because the orbit separation at the moment of
SN explosion from this channel is significantly less than that of
the WD + MS channel, which may result in a much stronger impact to
the companion. The surviving companion star from the WD + He star
channel could be verified by future observations.

\section{DISCUSSION AND CONCLUSION} \label{5:DISCUSSION AND CONCLUSIONS}
Wang et al. (2009), based on equation (1) of Iben $\&$ Tutukov
(1984), estimated the potential SN Ia birthrate through the WD + He
star channel to be $\sim 1.2\times 10^{-3}\,{\rm yr}^{-1}$ in the
Galaxy. The birthrate from Wang et al. (2009) is higher than that in
this paper, this is due to the fact that the long orbit period
(i.e., $\ga 1$ day in Fig. 8 of Wang et al. 2009) systems, which are
considered to produce SNe Ia in equation (1) of Iben \& Tutukov
(1984), do not contribute to SNe Ia in the simulation results of
this paper. In addition, Umeda et al. (1999a) concluded that the
upper limit mass of CO cores born in binaries is about
$1.07\,M_\odot$. If this value is adopted as the upper limit of the
CO WD, the birthrate of SNe Ia from this channel will decrease to be
$\sim 0.2\times10^{-3}\,{\rm yr}^{-1}$ in the Galaxy according to
our standard model.

In this paper we assume that all stars are in binaries and about
50\,percent of stellar systems have orbital periods less than
100\,yr. In fact, it is known not to be the case, and the binary
fractions may depend on metallicity, environment, spectral type,
etc. If we adopt 40\,percent of stellar systems have orbital periods
below 100\,yr by adjusting the parameters in equation (7), we
estimate that the birthrate of SNe Ia from this channel will
decrease to be $\sim 0.24\times10^{-3}\,{\rm yr}^{-1}$ for our
standard model.

SNe Ia from the WD + He star channel usually have massive CO WDs as
their progenitors. Some previous studies showed that a massive CO WD
leads to a lower C/O ratio in the Chandrasekhar-mass WD, and thus a
lower amount of $^{\rm 56}{\rm Ni}$ synthesized in the thermonuclear
explosion, which results in a lower luminosity of SNe Ia (Umeda et
al. 1999b; Nomoto et al. 1999, 2003). However, brighter SNe Ia more
frequently occur in active star formation galaxies (Hamuy et al.
1995, 1996), in which the young stellar population implies that
these SNe Ia have short delay times (see also Aubourg et al. 2008),
i.e., the CO WD + He star channel might produce brighter SNe Ia.
Therefore, it is difficult to explain SN Ia diversity by using the
C/O ratio. Note that 3D simulations about SN Ia explosions by
R\"{o}pke $\&$ Hillebrandt (2004) also indicated that different C/O
ratios have a negligible effect on the amount of $^{56}$Ni produced.
To understand the diversity of SN Ia explosions, the formation of
brighter SNe Ia should be explored in future investigations.

It is suggested that the WD + He star systems may appear as
supersoft X-ray sources (SSSs) prior to SN Ia explosions (Iben \&
Tutukov 1994; Yoon \& Langer 2003; Wang et al. 2009). Recently, Di
Stefano \& Kong (2003) used a set of conservative criteria,
applicable to {\em Chandra} data, to identify luminous SSSs in four
external galaxies (an elliptical galaxy, NGC 4967; two face-on
spiral galaxies, M101 and M83; and an interacting galaxy, M51). They
found that in every galaxy there are at least several hundred
luminous SSSs with a luminosity of $10^{37}\,{\rm erg\,s^{-1}}$, and
that in spiral galaxies M101, M83 and M51, the SSSs appear to be
associated with spiral arms. This may indicate that some SSSs are
young systems, possibly younger than $10^{8}$\,yr. Note that a WD +
He star system has X-ray luminosity around $10^{37}-10^{38}\,{\rm
erg\,s^{-1}}$ when He burning is stable on the surface of the WD
(Wang et al. 2009). Meanwhile, the distribution of SNe Ia with short
delay times associated with galactic spiral arms (Bartunov et al.
1994; Della Valle \& Livio 1994). Therefore, we emphasize that these
WD + He star systems may be related to the young SSSs prior to SN Ia
explosions.

The most important conclusion of this study is that the WD + He star
channel can explain SNe Ia with short delay times
($\sim4.5\times10^7 - 1.4\times10^8$\,yr), which is consistent with
recent observational implications of young populations of some SN Ia
progenitors (Scannapieco \& Bildsten 2005; Mannucci et al. 2006;
Aubourg et al. 2008). The young population of SNe Ia may have an
effect on models of galactic chemical evolution, since they would
return large amounts of iron to the interstellar medium much earlier
than previously thought. It may also have an impact on cosmology, as
they are used as cosmological distance indicators.

\acknowledgments We thank an anonymous referee for his/her valuable
comments that helped to improve the paper. BW thanks Dr. Richard
Pokorny for improving the English language of the original
manuscript. This work is supported by the National Natural Science
Foundation of China (Grant Nos. 10521001, 2007CB815406 and
10603013), the Foundation of the Chinese Academy of Sciences (Grant
No. 06YQ011001) and the Yunnan Natural Science Foundation (Grant No.
08YJ041001).



\label{lastpage}
\end{document}